\documentclass[prd,twocolumn,nofootinbib,aps,superscriptaddress,tightenlines,preprintnumbers]{revtex4}

\usepackage{epsfig}

  \newcount\hour \newcount\minute
  \hour=\time \divide \hour by 60
  \minute=\time
  \newcommand{\mydate}{\ \today \ - \number\hour :\ifnum \minute<10 0\fi 
\number\minute}
\begin{document}

\preprint{\hbox{CALT-68-2614}  }

\title{Minimal Extension of the Standard Model Scalar Sector}

\author{Donal O'Connell}
\affiliation{California Institute of Technology, Pasadena, CA 91125 }

\author{Michael J. Ramsey-Musolf}
\affiliation{California Institute of Technology, Pasadena, CA 91125 }
\affiliation{University of Wisconsin-Madison, Madison, WI 53706}

\author{Mark B. Wise}
\affiliation{California Institute of Technology, Pasadena, CA 91125 }

\begin{abstract}
The minimal extension of the scalar sector of the standard model contains
an additional real scalar field with no gauge quantum numbers. Such
a field does not couple to the quarks and leptons directly but rather
through its mixing with the standard model Higgs field. We examine the
phenomenology of this model focusing on the region of parameter space
where the new scalar particle is significantly lighter than the usual
Higgs scalar and has small mixing with it. In this region of parameter
space most of the properties of the additional scalar particle are
independent of the details of the scalar potential. Furthermore the
properties of the scalar that is mostly the standard model Higgs can be
drastically modified since its dominant branching ratio may be to a pair
of the new lighter scalars.

\end{abstract}

\date{\today}

\maketitle

%
%%%%%%%%%%%%%%%%%%%%%%%%%%%%%%%%%%%%%%%%%%%%%
\section{Introduction}
\label{sec:intro} 

Many of the extensions of the standard model that are testable at the
LHC have been motivated by the hierarchy puzzle. However, the small
value of the cosmological constant requires a fine tuning that in many
ways is quite similar to the fine tuning needed to keep the Higgs scalar
light. Most extensions of the standard model that are constructed to solve
the hierarchy puzzle still require a fine tuning to keep the cosmological
constant small. It is possible that we are not looking at this issue
correctly and that stability of radiatively corrected scalar potential
should not be the motivation for extensions of the standard model.

The scalar sector of the standard model has not been tested directly
by experiment, so it is certainly worth examining models with simple
extensions of this sector and exploring their phenomenology in some
detail. There does not seem to be a compelling motivation for a scalar
sector that consists of just a single Higgs doublet. However, if one
does not adopt additional symmetry principles~\cite{Glashow, Georgi} then
adding more doublets typically gives unacceptably large tree level flavor
changing neutral currents. Furthermore scalar fields with nontrivial
$SU(2)\times U(1)$ quantum numbers that are different from those of the
standard model Higgs doublet must have small vacuum expectation values to
preserve the standard model value of the $\rho$ parameter. The phenomenology
of the standard model Higgs scalar, models with multiple Higgs doublets,
and of supersymmetric extensions of the standard model has been studied
extensively, and Ref.~\cite{cahn} contains some excellent reviews.

The simplest extension of the scalar sector of the minimal standard
model is to add a single real scalar $S$ that is a gauge singlet. This
does not have to be a fundamental degree of freedom; the Higgs doublet
and this scalar might be the only light remnants of a more complicated
scalar sector that manifests itself at scales that are too high to be
directly probed by the next generation of accelerator experiments. In
this paper we examine the phenomenology of this extension of the standard
model. Extensions of the minimal standard model with one or more singlets
$S$ have been studied before in the literature. Many of the models impose
a $S \rightarrow -S$ symmetry, so that the singlet can be a dark matter
candidate. Other works (for example, see Ref.~\cite{Krasnikov}) either
do not impose a $S \rightarrow -S$ symmetry or break that symmetry, but
have some differences with the model we present here (e.g., some possible
couplings in the scalar potential are missing, or the lighter scalar is
taken to be massless, etc.) In any case, it seems worth reexamining the
phenomenology of this model since we are approaching the LHC era. Our work
was inspired by Ref.~\cite{Ellis} where an additional scalar superfield
was added to the minimal supersymmetric standard model to solve the $\mu$
problem, and some of our conclusions are similar to theirs.

In the minimal standard model the scalar potential for the Higgs doublet
$H$ contains only two parameters which can be eliminated in favor of
the Higgs particle mass and the vacuum expectation value that breaks
$SU(2)\times U(1)$ gauge symmetry. When the singlet scalar $S$ is added
the number of parameters of the scalar potential swells to seven. However
for most of the phenomenology only a few parameters are relevant, and a
very simple picture emerges. The singlet scalar $S$ and the Higgs scalar
$h$ mix, and both of the resulting physical particles have couplings to
quarks, leptons and to gauge bosons that are proportional to those of
the standard model Higgs particle. In addition to decays to the standard
model fermions and gauge bosons, the heavier of these two scalar particles
may decay to a pair of the lighter ones.

In this brief report we focus on the region of parameter space where the
lighter of the two scalar particles is mostly singlet and has a small
enough mass so that it can be pair produced in decays of the heavier
(mostly) Higgs scalar. This will be the most interesting case for LHC
physics. The only new parameters (beyond those in the standard model)
that are needed to characterize most of the phenomenology of this model
are the $h-S$ mixing angle, the mass of the new light scalar particle,
and the branching ratio for the decay of the heavier Higgs scalar to a
pair of the lighter ones.

\section{Scalar Potential}
\label{sec:pot} 
The Lagrange density for the scalar sector of this model is
\begin{equation}
{\cal L}=(D_{\mu}H)^{\dagger}D^{\mu}H+{1 \over 2}\partial_{\mu}S\partial^{\mu}S -V(H,S),
\end{equation}
where $H$ denotes the complex Higgs doublet and $S$ the real scalar. Without loss of generality, we shift the field $S$ so it has no vacuum expectation value. Then the potential is given by
\begin{eqnarray}
&&V(H,S)={m^2 \over 2} H^{\dagger}H +{\lambda \over 4} (H^{\dagger}H)^2 +{\delta_1 \over 2}H^{\dagger}H~S  \\
&&+{\delta_2 \over 2}H^{\dagger}H~S^2+\left( {\delta_1 m^2 \over 2 \lambda}\right)S+{\kappa_2 \over 2} S^2 +{\kappa_3 \over 3}S^3+{\kappa_4 \over 4}S^4. \nonumber 
\end{eqnarray}
Note that there is no additional $CP$ violation that comes from the scalar potential.

In unitary gauge the charged component of the Higgs doublet $H$ becomes the longitudinal components of the charged $W$-bosons and the imaginary part of the neutral component becomes the longitudinal component of the $Z$-boson. The neutral component is written as 
\begin{equation}
H^0= {v+h \over \sqrt{2}},~~~~~ v=\sqrt{-2m^2 \over  \lambda}.
\end{equation}
The mass terms in the scalar potential become
\begin{equation}
V_{ \rm mass}={1 \over 2}\left(\mu_h^2 h^2 +\mu_S^2 S^2 +\mu_{hS}^2 hS \right),
\end{equation}
where
\begin{eqnarray}
\label{param}
&&\mu_h^2=-m^2=\lambda v^2/2 \nonumber \\
&&\mu_S^2= \kappa_2+\delta_2 v^2/2  \nonumber \\
&&\mu_{hS}^2=\delta_1 v.
\end{eqnarray}
The mass eigenstate fields $h_+$ and $h_-$ are linear combinations of the Higgs scalar field $h$ and the singlet scalar field $S$. Explicitly, for the lighter field $h_{-}$
\begin{equation}
h_-= {\rm cos}\theta ~S- {\rm sin}\theta ~h,
\end{equation}
where~\cite{Krasnikov}
\begin{equation}
\tan \theta={ x \over 1+\sqrt{1+x^2}},~~~x={\mu_{hS}^2 \over \mu_h^2 -\mu_S^2}.
\end{equation}
The terms in the scalar potential that break the discrete $S \rightarrow
-S$ symmetry are proportional to the couplings $\delta_1$ and $\kappa_3$
so it is natural for those scalar couplings to be small. The parameter
$\delta_1$ controls the mixing of the two scalar states. As we will see, 
experimental constraints force the mixing angle $\theta$ to be small.
% and we work in the limit of small mixing angle $\theta$. 
We assume the heavier
state is mostly the Higgs scalar so $\mu_h^2 > \mu_S^2$. The masses of
the two scalars are
\begin{equation}
m_{\pm}^2=\left({\mu_h^2 +\mu_S^2 \over 2 }\right)\pm \left({\mu_h^2 - \mu_S^2 \over 2 }\right) {\sqrt{1 +x^2}}.
\end{equation}

\section{Phenomenology}
The lighter scalar state decays to standard model particles and
its couplings to them are proportional to the standard model Higgs
couplings with constant of proportionality $\sin\theta $. Consequently,
the lighter of the two states has branching ratios equal to those of the
standard model Higgs (if it had mass $m_-$) and its production rates are
$\sin^2\theta $ times the production rates for a standard model Higgs (if
it had mass $m_-$). Since $\sin \theta $ can be much smaller than unity,
$m_-$ can be much smaller than the mass of the standard model Higgs,
which is restricted by the LEP bound to be heavier than $114~{\rm GeV}$.

If $m_+ <2m_-$ then the heavier state has branching ratios to standard model particles and production rates approximately equal to those of the standard model Higgs scalar (recall we are working in the limit of small mixing). However if $m_+>2m_-$ then the decay channel $h_+ \rightarrow h_- h_-$ is available with partial decay width
\begin{equation}
\Gamma(h_+\rightarrow h_-h_-)={\delta_2^2v^2 \over 32 \pi m_+ } \sqrt{1-4m_-^2 /m_+^2} .
\end{equation}
For a $h_+$ that has mass below $140{\rm GeV}$ its dominant decay mode to standard model particles is to a bottom anti-bottom pair and so in this mass range
\begin{equation}
{\Gamma(h_+\rightarrow h_-h_-)\over \Gamma^{\rm S.M.}(h)}\simeq {\delta_2^2 v^4 \over 6 m_+^2 m_b^2 } \sqrt{1-4m_-^2 /m_+^2},
\end{equation}
where $\Gamma^{\rm S.M.}(h)$ denotes the decay width of the standard model Higgs. Conventional branching ratios of the heavier scalar $h_+$ are reduced from those of the standard model Higgs by a factor $f$ which is equal to
\begin{equation}
f={1 \over 1+\Gamma(h_+\rightarrow h_-h_-)/\Gamma^{\rm S.M.}(h)}=1-{\rm Br}(h_+\rightarrow h_-h_-).
\end{equation}

\subsection{Very Light $h_-$}
If $m_-$ is much smaller than the weak scale it seems most natural to
take $|\delta_2|$ to be of order $2m_-^2/v^2$ or smaller, since if it
is much larger than this a strong cancellation between the two terms
contributing to $\mu_S^2$ in Eq.~(\ref{param}) is necessary. Suppose
$m_-$ is less than the $b$-quark mass, $m_b$. With $|\delta_2| \sim 2
m_-^2/v^2$ we find that in this region of parameter space $f$ is very
close to unity since $\Gamma(h_+\rightarrow h_-h_-)/\Gamma^{\rm S.M.}(h)
\sim (m_-/m_b)^2(m_-/m_+)^2 \ll 1$. The heavier scalar state will behave
much like the standard model Higgs particle.

The properties of a very light $h_-$ are severely constrained by present
experimental data. Consider, for example, the case where $m_-=500~{\rm
MeV}$. Then the branching ratio of the $h_+$ to two $h_-$'s is very
small and the $h_+$ is indistinguishable from the standard model
Higgs particle. The dominant decay modes of the $h_-$ are to two
pions and to $\mu^+ \mu^-$ with partial decay rates\footnote{The rate
to two pions is calculated at leading order in chiral perturbation
theory~\cite{chiral}. With $m_-=500~{\rm MeV}$ one can expect sizeable
corrections from, for example, $\pi \pi$ final state interactions. These
are expected to increase the decay rate to two pions.},
\begin{eqnarray}
&&\Gamma(h_- \rightarrow \pi^+\pi^-)=2\Gamma(h_- \rightarrow \pi^0\pi^0)={{\rm sin}^2\theta~m_-^3 \over 324 v^2\pi} \nonumber \\
&&\times\left[1+{11 \over 2}\left({m_{\pi}^2 \over m_-^2}\right)\right]^2 \sqrt{1-{4m_{\pi}^2 \over m_-^2}},
\end {eqnarray}
and
\begin{equation}
\Gamma(h_- \rightarrow \mu^+\mu^-)={{\rm sin}^2\theta~m_- m_{\mu}^2\over 8 v^2\pi}\left[1-{4m_{\mu}^2 \over m_-^2}\right]^{3/2}.
\end{equation}
These imply that ${\rm Br}(h_ -\rightarrow \mu^+ \mu^-)\simeq 34\%$
and that the lifetime of the $h_-$ is $\tau_{h_-}\simeq (9/{\rm
sin}^2\theta)\times 10^{-17}~\rm{sec}$. The strongest constraint on
the mixing angle $\theta$ comes from the decay $B \rightarrow h_- X $
which has branching ratio ${\rm Br}(B \rightarrow h_- X) \simeq 8 ~{\rm
sin}^2\theta$. The experimental limit ${\rm Br}(B \rightarrow \mu^+\mu^-
X)<3.2~\times 10^{-4}$~\cite{pdg} implies that ${\rm sin}^2\theta <
1 \times 10^{-4}$. Hence the lifetime of the $h_-$ is at least $8\times
10^{-13} {\rm sec}$ (i.e., about the $B$ meson lifetime).

The $h_-$ can be produced directly in $Z$ decays. Single $h_-$ production
is suppressed by the the small mixing angle, ${\rm Br}(Z \rightarrow h_-
\bar f f )/{\rm Br}(Z\rightarrow \bar f f) \simeq 10^{-2}{\rm sin}^2\theta
< 1 \times 10^{-6}$, where $f$ denotes any of the light standard model
fermions. The $h_-$ can also be pair-produced through a virtual $h_+$,
with a rate that is not suppressed by the small mixing angle $\theta$, via
the process $Z \rightarrow h_+^* \bar f f \rightarrow h_-h_- \bar f f$. We
find that this rate is negligible\footnote{For $\kappa_2=0$, $m_+=120~{\rm
GeV}$ and $m_-=5{\rm GeV}$, we find that ${\rm Br}(Z \rightarrow h_+^*
\bar f f \rightarrow h_-h_- \bar f f)/{\rm Br}(Z\rightarrow \bar f
f)=5.4\times 10^{-13}$. The rate is even smaller for smaller values of
$m_-$.} when $|\delta_2| \sim 2 m_-^2/v^2$.

\subsection{$5~{\rm GeV}< m_- <50~{\rm GeV}$}

This mass range is interesting because the decays of the heavier scalar,
which is mostly the standard model Higgs, can be quite different
from what the minimal standard model predicts. In this mass range
the $h_-$ is light enough that the decay $h_+ \rightarrow h_-h_-$
is kinematically allowed. In addition the $h_-$ is heavy enough so that values of $\delta_2$
that give this decay a significant branching ratio
do not require a delicate cancellation between the two terms contributing to $\mu_S^2$ in Eq.~(\ref{param}). At the lower end of this mass range we
would not expect the dominant decay of the $h_+$ to be to two $h_-$'s;
however, with no awkward cancellation between the two terms contributing
to $\mu_S^2$, the $h_+$ could easily decay about the same amount of time
to two $h_-$'s as it does to two photons. At the upper end of the mass range it is quite reasonable to
have the $h_+$ decaying mostly to two $h_-$'s.

\begin{figure}[t]
\includegraphics[width=0.5\textwidth]{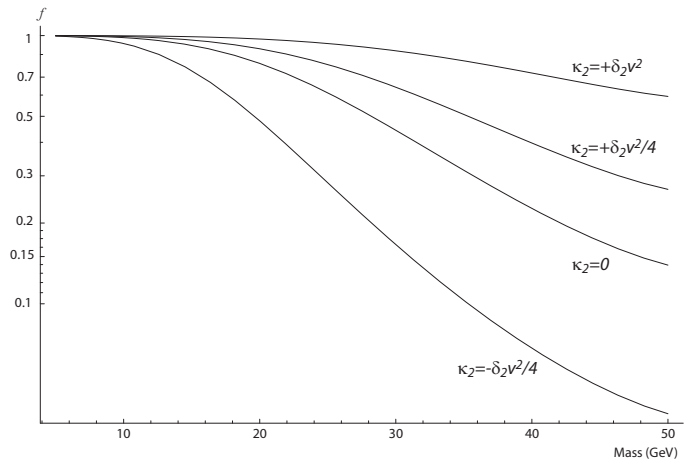}
\caption{The suppression factor $f$ discussed in the text.}  
\label{fig:suppression}
\end{figure}

In Fig.~\ref{fig:suppression} we plot the suppression factor $f =1-{\rm
Br}(h_+\rightarrow h_- h_-) = {\rm Br}(h_+\rightarrow\gamma\gamma)/{\rm
Br}^{\rm S.M.}(h\rightarrow\gamma\gamma)$, where ${\rm Br}^{\rm S.M.}(h
\rightarrow \gamma \gamma)$ is the standard model branching fraction,
assuming that the parameter $\kappa_2=-\delta_2 v^2/4,0, +\delta_2 v^2/4,
+\delta_2 v^2$ and that the mixing angle $\theta$ is very small. Notice
that as one approaches the upper range of the mass range we are
considering, decays of the $h_+$ can be dominated by the final state
$h_-h_-$, and consequently the branching ratio to the the two photon mode
is suppressed compared to what it is in the standard model. The dependence
of $f$ on the mass of the $h_-$ arises because the same coupling in the
Lagrangian that gives rise to this branching ratio also contributes to
the $h_-$ mass.  If $\kappa_2$ is larger than $\delta_2 v^2$, then $f$
will be close to unity throughout the whole range we consider.

\begin{figure}[b]
\centering
\includegraphics[width=0.5\textwidth]{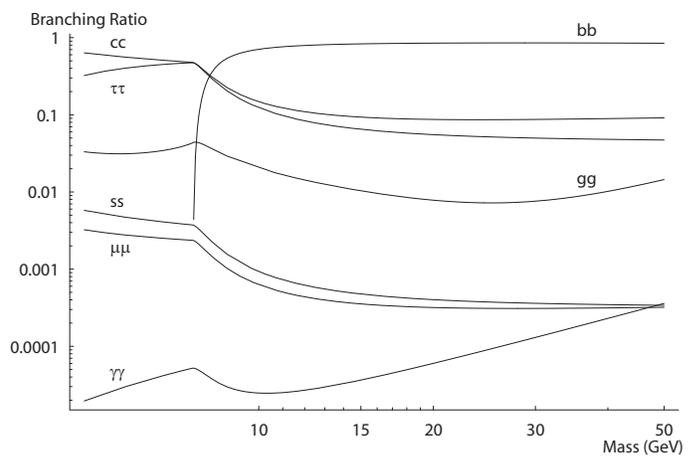}
\caption{The branching ratios of the light $h_-$ scalar particle.}
\label{fig:br}
\end{figure}

Since the $h_+$ may decay dominantly to a pair of the lighter $h_-$
scalars it is useful to understand the ensuing $h_-$ decays. We plot
the branching ratios of the $h_-$ in Fig.~\ref{fig:br}. In the upper
end of the mass range we are considering, the $h_-$ decays mostly to a
bottom-antibottom pair but it also has substantial branching fractions
to $c\bar c$ and to $\tau^+ \tau^-$. Considerable attention has
been given to a different class of models where the the heavier Higgs
scalar decays mostly to a singlet scalar that consequently decays to
neutrinos~\cite{invisible}.

In Ref.~\cite{Manohar} it was noted that new physics at the ${\rm TeV}$ scale can easily give rise to a large reduction (or enhancement) in the dominant gluon fusion Higgs scalar production rate, and hence the final two photon signal. However, it is very unlikely that such physics could alter the associated production rate of the Higgs since it arises from the tree coupling to the massive weak bosons. The situation is different here. All the standard model decays of the heavier Higgs-like scalar $h_+$ are reduced by the same factor $f$ independent of its production mechanism.

The production of $h_-$ scalars from gluon fusion is suppressed from
the production rate for a standard model Higgs of the same mass by
${\rm sin}^2\theta$. Since the $h_-$ was not observed at ${\rm LEP}$
there is a mass dependent limit on ${\rm sin}^2\theta $ from ${\rm LEP}$
data~\cite{Barate:2003sz} (Very roughly, ${\rm sin}^2\theta <2\times
10^{-2}$ over the mass range we are considering.). However, the $h_-$
production rate via gluon fusion at the Tevatron and LHC increases rapidly
as its mass decreases. For example, using leading order CTEQ5 parton
distributions~\cite{cteq} an $h_-$ of mass $10~{\rm GeV}$ has a production
rate roughly $100$ times greater than one with a mass of $120~{\rm GeV}$
at the LHC and $1,000$ times greater at the Tevatron. Even with a 
small value for ${\rm sin}^2\theta$ the $h_-$ may be directly observable
at the LHC.

\section{Concluding Remarks}

The literature on extensions of the minimal standard model with a more
complicated scalar sector is vast. Here, we have considered the simplest
alternative possibility, in which a single gauge-singlet scalar is
added. It mixes with, and couples to, the standard model Higgs. We
concentrate on the region of allowed singlet scalar masses where it
is light enough to be pair produced in Higgs decay and, hence, suppress
the Higgs \lq\lq golden mode" branching ratio to two photons. We found
that this suppression is unlikely to be significant if the new scalar
is very light but can easily be large if the (mostly) singlet scalar
is heavier than about $10~{\rm GeV}$. Under this scenario, there is
a scalar that has a mass below the LEP Higgs mass lower bound with decay
branching ratios that are identical to those of a standard model Higgs
of the same mass. However its production rate is suppressed by a small
mixing angle. Potential signatures of such light scalars include the
observation of Higgs decay products with invariant mass well below 114
GeV or unusual final states in Higgs decay such as two $b$-jets and a
$\tau^+ \tau^-$ (or $\mu^+ \mu^-$) pair.

%If observed, the presence of such singlet scalars could also have
%implications for cosmology. While the small, non-vanishing Higgs-singlet
%mixing assumed in our analysis would preclude a relic density of scalar
%singlet dark matter, it could provide one of the necessary ingredients
%for production of the baryon asymmetry at the electroweak scale. In the
%absence of cancellations among parameters in $V(H,S)$ (prior to shifting
%the field $S$ to remove its vev), the cubic $H^\dag H S$ coupling would
%be large enough to provide for the requisite strong first order phase
%transition if $\sin\theta\gsim 0.1$. This requirement, which is roughly
%independent of the SM-like Higgs mass and is not precluded by LEP limits,
%could in principle be tested by detailed studies of light scalars at
%the Tevatron and LHC.

\end{document}